\documentclass[aps,prd,twocolumn,groupedaddress,showpacs,showkeys,nofootinbib,amssymb]{revtex4-1}
\usepackage[latin1]{inputenc}
\usepackage{graphicx}
\usepackage[english]{babel}
\usepackage{amsmath}
\usepackage{amssymb}
\usepackage{amsfonts}
\usepackage{colordvi}
\usepackage{psfrag}
\usepackage{color}
\usepackage{tipa}
\begin{document}
\def\pp{{\, \mid \hskip -1.5mm =}}
\def\cL{{\cal L}}
\def\be{\begin{equation}}
\def\ee{\end{equation}}
\def\bea{\begin{eqnarray}}
\def\eea{\end{eqnarray}}
\def\beq{\begin{eqnarray}}
\def\eeq{\end{eqnarray}}
\def\tr{{\rm tr}\, }
\def\nn{\nonumber \\}
\def\e{{\rm e}}

%\title{\textbf{Scaling astrophysical structures from Eddington-Weinberg relations}}
\title{\textbf{Primordial black holes, astrophysical systems  and the Eddington-Weinberg relation}}

\author{Salvatore Capozziello$^{1,2}$, Gerardo Cristofano$^{1,2}$, Mariafelicia  De Laurentis$^{1,2}$}

\affiliation{\it $^{1}$Dipartimento di Scienze Fisiche, Università
di Napoli {}``Federico II'',\\
 $^{2}$INFN Sez. di Napoli, Compl. Univ. di
Monte S. Angelo, Edificio G, Via Cinthia, I-80126, Napoli, Italy}

\date{\today}
\begin{abstract}
Starting from a quantization relation for primordial black holes,
it is shown that quantum fluctuations can play a fundamental role in determining the effective scales of self-gravitating astrophysical systems. Furthermore the  
Eddington-Weinberg relation between the current scale of the observed universe to the  Planck constant (the natural action unit)  is naturally derived. Finally, such an approach allows to recover  the current value of the cosmological constant.

 \end{abstract}
 \keywords{ cosmology; quantum field theory; large scale structure; cosmological constant }
\pacs{04.50.+h, 95.36.+x, 98.80.-k}

\maketitle

\section{Introduction}
\label{uno}

The theoretical understanding of the observed scales, sizes and dimensions of self-gravitating aggregated structures in the universe (stars, galaxies, clusters, etc.) is a long-standing open problem in astrophysics and cosmology.
Together with this issue, there have been many attempts to connect such macroscopic features with primordial quantities  at the  very beginning of the universe history, that is at Planck epoch and beyond. 
Besides,  several decades ago Dirac and Eddington found certain coincidences between large numbers, relating the size and the age of the observed universe to the   Compton wavelength of the proton   and to the time for traveling the proton size at light speed, respectively \cite{dirac1,dirac2,dirac3,dirac4,eddington}. Furthermore Dirac strongly believed that such relations are not  mere coincidences but  hints for some fundamental laws of nature.  
In general, such relations appear very interesting since they  connect macroscopic scales, as for example the present radius of the universe, with the quantum constant
 $\hbar$ which can be considered as the natural action unit  \cite{weinberg1,weinberg2,demartino}. In other words, one would be induced to think that all macroscopic structures (and the universe itself)  have taken origin from quantum phenomena (quantum fluctuations), which were at work at the origin of the universe. In some sense, this is the philosophy underlying the  Quantum Cosmology \cite{halliwell}.
In the context of large numbers coincidences, also the cosmological constant $\Lambda$, today considered the main ingredient  for the universe acceleration, can be  expressed in terms of the  Compton wavelength of the proton,  even though its previous estimate by Zeldovich is off of several order of magnitude with respect to the  current estimates \cite{zeldovich}.
Unfortunately, at the moment, there is no non-perturbative field theory  capable of  giving such coincidences as a self-consistent finding  or  establishing any definite relation between large and small scales. 

In this  paper, we propose a straightforward approach to this problem, based essentially on quantum mechanics. Our starting point is the Dirac quantization relation applied to charged primordial black holes \cite{SGF}.
The main  result of such a description relies on the scaling properties of  quantum relations that can be directly connected to astrophysical systems and large scale structure.
The paper is organized as follows. Sec. II is devoted to the scaling properties of  quantization relation for primordial black holes which can be considered as the seeds of large scale structure.
In Sec. III,  the effective  scales of self-gravitating astrophysical  structures are derived by using the quantization relation for black holes. A fundamental constant   $a_0$ with the dimensions of an acceleration is derived. 
In Sec. \ref{quattro},  the Eddington-Weinberg relation is reproduced while, in Sec. \ref{cinque},  it is shown how  the  current value of the cosmological constant $\Lambda$ is recovered. Conclusions are drawn in Sec. \ref{sei}.

%%%%%%%%%%%%%%%%%%%%%%%%%%%%%%%%%%
\section{Scaling hypothesis from black holes quantization}
\label{due}
%%%%%%%%%%%%%%%%%%%%%%%%%%%%%%%%%%%%%%%%

In a previous paper \cite{SGF}, a quantization relation has been  proposed for black holes  by adopting the Dirac quantization condition that relates  the electric and magnetic charges of a black hole to its mass. It is interesting to stress that such a quantization, derived at Planck scales,  can be the seed of  scaling relations for  all self-gravitating astrophysical systems that we  observe now in the universe. Here we explicitly give such a connection,  emphasizing the role played by quantum fluctuations at the very early epochs.
Let us shortly review the quantization relation derived in \cite{SGF}, starting from the black hole effective potential
\begin{equation}
V_{BH}=e^{2\phi}\left(Q_e-aQ_m\right)^2+e^{-2\phi}Q^{2}_{m}\,,
\label{1}
\end{equation}
where $Q_e$ and $Q_m$ are the electric and magnetic charges of the black hole, $\phi$ is the dilaton field and "$a$" the axion field. We recall that this kind of  black holes are called {\it  extremal}, that is their event horizons become degenerate.
Besides,  from Conformal Field Theory, we have
\begin{equation}
V_{eff}^{CFT}=R_{c}^{2}\left(Q_e-\frac{\theta}{2\pi}Q_m\right)^2+\frac{1}{R_{c}^{2}}Q^{2}_{m}\,,
\label{2}
\end{equation}
where $R_c$ is the compactification radius of the so-called  {\it Fubini scalar  field} \cite{fubini} and $\displaystyle{\frac{\theta}{2\pi}}$ is the  {\it theta parameter} 
\cite{Gthooft}.
Comparing Eq. (\ref{1}) with Eq. (\ref{2}), we get the following identifications
\begin{equation}
R_{c}^{2}=e^{2\phi}\,, \qquad \frac{\theta}{2\pi}=a\,.
\label{3}
\end{equation}
Without loosing  generality, we will consider the case  $\displaystyle{a=\frac{\theta}{2\pi}=0}$  from now on, that is we will discuss a black hole effective potential  in which only the dilaton field is present \cite{kallosh}:

\begin{equation}
V_{BH}=e^{2\phi}Q_e^2+e^{-2\phi}Q^{2}_{m}\,.
\label{1a}
\end{equation}
Using such a potential and
 imposing the criticality condition
\begin{equation}
\frac{\partial V_{BH}}{\partial\phi}=0\,,
\label{4}
\end{equation}
one obtains
\begin{equation}
e^{2\phi_{H}}=\frac{Q_m}{Q_e}=R_{H}^{2}
\label{5}
\end{equation}
where
$\phi_{H}$ and $R_{H}$ indicate the corresponding values at the horizon of the black hole. Furthermore since such black holes are also extremal, their mass saturates the so-called "Bogomol'nyi-Prasad-Sommerfeld"   bound (e.g., see \cite{Ferrara})
\begin{equation}
M^2=e^{2\phi_{H}}Q_{e}^2+e^{-2\phi_{H}}Q^{2}_{m}\,.
\label{6}
\end{equation}
Notice that, in the metric approach,  Eq.(\ref{6}) is a consequence of assuming the metric time-time component $g_{tt}=0$. This is an indication of a phase transition, that is charged  black holes are forming with mass $M$ and charges $Q_e$, $Q_m$ obeying Eq.(\ref{6}) (see \cite{SGF}). Furthermore, substituting Eq.(\ref{5}) into Eq.(\ref{6}), we obtain
\begin{equation}
M^2=2Q_e Q_m\,.
\label{7}
\end{equation}
In order to consider black holes as quantum objects, we assume the  Dirac quantization condition
\begin{equation}
2Q_e Q_m=n\hbar c\,,
\label{8}
\end{equation}
with $n$ a positive integer \cite{SGF}. Substituting into the previous relation and introducing standard units, one gets
\begin{equation}
GM^2 =n\hbar c\,.
\label{9}
\end{equation}
For $n=1$, we obtain the lowest mass allowed for a quantum black hole (primordial black hole):
\begin{equation}
M_{BH}=\sqrt{\frac{\hbar c}{G}}=M_{Planck}\,.
\label{10}
\end{equation}
It is interesting to notice that, in the quantum relation (\ref{9}), there is no remnant of the electric and magnetic charges of the object of mass $M$, instead  its angular momentum $J = n\hbar$ appears on the right side of the relation thanks to Dirac quantization. The relation (\ref{9}) has been proved  to be valid for any "self-gravitating" systems  \cite{SGF}.

Here we want  to study the scaling properties of this quantum relation. To this end,  let us recall that the factor $n$ can be expressed, considering the Compton length 
$\lambda$ and the Schwarzschild radius  as
\begin{equation}
n_{as}=\frac{R_{as}^{Schw}}{2\lambda_{as}}=\left(\frac{GM_{as}}{c^2}\right)\frac{n^{as}_{p}}{\lambda_{p}}\,,
\label{11}
\end{equation}
where the label {\it "as"} refers to the generic  astrophysical structure considered and $n_{p}^{as}$ is the number of protons contained in it. 

This is the key ingredient of our approach that has to be discussed in details. Eq.(\ref{11}) points out that the characteristic length of gravitational interaction, the Schwarzschild radius, and the  characteristic quantum length for the same structure are related by the granular components of the structure itself, the protons. Such a relation is not arbitrarily assumed but ruled by the Planck length and mass of primordial black holes. Such primordial structures  are the first quantized structures emerging from early phase transitions  \cite{Ferrara} and, according to several inflationary cosmological models, can be considered the seeds for large-scale structure. It is worth stressing that relation (\ref{11}) strictly depends on the quantum and gravitational interactions and it is not a mere "close packing" where astrophysical and cosmological structures can be built up by the sum of granular elements (see also \cite{demartino,SGF,calogero}).  In this sense, it is not only the "number" of protons that constitute the structure but mainly their mutual interactions. 

With these considerations in mind, Eq.(\ref{11}) can be recast as 
\begin{equation}
n_{as}=const\cdot (n_{p}^{as})^2\,,
\label{12}
\end{equation}
and  specializing Eq. (\ref{11}) to the case of the universe \cite{eddington,weinberg2}, it is not difficult to find that
\begin{equation}
const=\frac{1}{(n_{p}^{BH})^2}\,,
\label{13}
\end{equation}
where $n_{p}^{BH}$ indicates the number of protons in the lowest  black hole of mass given by Eq.(\ref{10}). 
Such a quantity is a constant since, depending on Eq.(\ref{10}), it is built up only by fundamental constants.
The basic quantization  relation (\ref{9}) can be written in a more useful form as
\begin{equation}
GM_{as}^{2}=\left(\frac{n_{p}^{as}}{n_{p}^{BH}}\right)^2\hbar c\,.
\label{14}
\end{equation}
 Finally using for the mass of a given astrophysical structure $M{_{as}}=n^{as}_{p}M_{p}$, with $M_{p}$ the proton mass, we obtain
 \begin{equation}
 G(n_{p}^{as}M_{p})^{2}=\left(\frac{n_{p}^{as}}{n_{p}^{BH}}\right)^2\hbar c\,.
 \end{equation}
 Furthermore multiplying both members by the factor $\displaystyle{\left(\frac{n_{p}^{BH}}{n_{p}^{as}}\right)^2}$, we get
\begin{equation}
G\left( n_{p}^{BH}M_{p}\right)^2=\hbar c\,,
\end{equation}
that is exactly
\begin{equation}
G M_{Planck}^{2}=\hbar c\,,
\end{equation}
We then conclude that the quantization relation (\ref{9}), or (\ref{14}),  scales  from the Planck mass to the mass of the different astrophysical structures up to the whole universe.

%%%%%%%%%%%%%%%%%%%%%%%%%%%%%%
\section{The physical size of self-gravitating  structures from the quantization relation}
\label{tre}
%%%%%%%%%%%%%%%%%%%%%%%%%%%%%%%%%%%%%%%%%

Let us now  show how the physical size of self-gravitating astrophysical systems can naturally arise  from the basic quantization relation (\ref{9}) or, equivalently, from (\ref{14}).
However, in this simplified model, we are assuming that gravity is the leading interaction and protons are the basic constituents.
Let us take into account the following relation 
\begin{equation}
 \frac{GM_{as}}{R^{2}_{as}}=\frac{GM_{u}}{R^{2}_{u}} =\frac{c^2}{R_u}=a_0\,.
\label{15}
\end{equation}
derived considering the Schwarzschild radius and an overall gravitational interaction acting up to the radius of the universe $R_u$ \cite{demartino,calogero}.
The constant $a_0$ has the physical dimensions of  an acceleration and  could have a much deeper meaning in connection to the structure formation \cite{Tula}. 
It will be discussed in detail elsewhere.

Notice that the previous equation implies
\begin{equation}
\frac{M_{as}}{R^{2}_{as}}=\frac{M_{u}}{R^{2}_{u}}=const\,,
\label{16}
\end{equation}
that we will use below.
The numerical factor $(n_{p}^{BH})^{-2}$ appearing in Eq. (\ref{14}), can be written in a more convenient form as
\begin{equation}
\frac{1}{(n_{p}^{BH})^2}=\left(\frac{M_{p}}{M_{BH}}\right)^2\simeq10^2\times 10^{-40}\simeq\frac{1}{\sqrt{n_{p}^{u}}}\,,
\label{17}
\end{equation}
where $n_{p}^{u}\simeq10^{80}$ indicates the number of protons in the universe.
Substituting it back into Eq. (\ref{14}), we  get
\begin{equation}
GM_{as}^2=10\frac{\left(n_{p}^{as}\right)^{2}\hbar c}{\sqrt{n_{p}^{u}}}\,.
\label{18}
\end{equation}
Assuming the relation (\ref{16}),  we obtain 
\begin{equation}
\sqrt{\frac{n_{p}^{as}}{n_{p}^{u}}}=\sqrt{\frac{n_{p}^{as}M_p}{n_{p}^{u}M_p}}=\sqrt{\frac{M_{as}}{M_u}}=\frac{R_{as}}{R_u}\,,
\label{19}
\end{equation}
and,  after dividing both members by $n_{p}^{as}M_{p}c^2$, we have
\begin{equation}
\frac{ G (n_{p}^{as}M_{p})}{c^2}=10\sqrt{n_{p}^{as}}\left(\frac{R_{as}}{R_u}\right)\left(\frac{h}{M_{p}c}\right)\,.
\end{equation}
Finally dividing left and right sides by $R^{2}_{as}$ and making use first of Eq. (\ref{16}) and then of Eq. (\ref{15}), we get
\begin{equation}
R_{as}\simeq 10 \sqrt{n_{p}^{as}} \lambda_{p}\,,
\label{20}
\end{equation}
which, fully reproduces the statistical hypothesis results (see \cite{scott1,scott2}) but with an important extra numerical factor of order $10$.
Specializing it to the case of the universe, we get
\begin{equation}
R_{u}\simeq 10 \sqrt{n_{p}^{u}} \lambda_{p}\simeq 10^{28}cm\,,
\label{21}
\end{equation}
In full agreement with current estimates of the observed radius of the universe. It is interesting to notice that the extra factors $10$ or $10^2$ are perfectly compatible with uncertainties in observations and could be related to the "scatter" between measurements of dark  and luminous matter.

An important issue has to be discussed at this point. Clearly we have not used any dark matter hypothesis and the characteristic sizes of astrophysical structures (in particular that of the observed universe) can be reproduced starting from primordial black holes and protons, assumed as granular elements. This is not surprising due to the fact that, in our model, the  Compton wavelength (the quantum interaction length) and the Schwarzschild radius (the gravitational interaction length) rule the astrophysical structures. From this point of view,  astrophysical structures are the macroscopic results of average stochastic processes which select their characteristic sizes (e.g. see the  book by Roy \cite{roy}). Finally, dark matter is not necessary in this framework since its effect (in particular "the missing matter issue") is addressed considering the gravitational interaction depending on scale. This is the viewpoint  of several alternative theories of gravity which  address the dark energy and dark matter problems without assuming new "dark" ingredients (up to now not detected at fundamental level) but shift the problem to the gravitational sector
\cite{farabook}.

%%%%%%%%%%%%%%%%%%%%%%%%%%%%%%%%%%%%%%%%%%%%%%%
\section{Reproducing  the Eddington-Weinberg relation}
\label{quattro}
%%%%%%%%%%%%%%%%%%%%%%%%%%%%%%%%%%%%%%%%%%%%%%%%

The above considerations can be used to reproduce some relations connecting quantum and cosmological scales.
In particular, the  famous Eddington-Weinberg relation  brings together the quantum unit of action $\hbar$ with the radius of the universe. 
It is interesting to derive this relation within our approach  which, as we have seen,  is based  on such a  link.
In order to do so, let us start from Eq. (\ref{20}), specializing it to the case of the universe, that is Eq.(\ref{21}).
Such a relation can be rewritten as
\begin{equation}
R_{u}\simeq 10 \sqrt{\frac{M_u}{M_p}}\left(\frac{h}{M_{p} c}\right),
\label{22}
\end{equation}
where the Compton length has been explicitly given. 
Writing it in a more convenient form
\begin{equation}
\frac{R_{u}}{\sqrt{M_u}}=10\frac{h}{M_{p}^{\frac{3}{2}}c}\,,
\end{equation}
and using $\displaystyle{\frac{R_{u}}{\sqrt{M_u}}=\frac{\sqrt{G R_u}}{c}}$ from Eq. (\ref{15}), we get, after substituting and rearranging terms,
\begin{equation}
h=\frac{1}{10}\sqrt{G R_u M_{p}^{3}}
\label{23}
\end{equation}
which fully reproduces Eddington-Weinberg relation (apart from the correcting factor $\frac{1}{10}$).
As a consistency check, the value of the radius of the universe can be obtained from Eq. (\ref{23}). It is 
\begin{equation}
R_u\simeq 10^{28}cm\,,
\end{equation}
in agreement with the corresponding value given in Eq. (\ref{21}).

%%%%%%%%%%%%%%%%%%%%%%%%%%%%%%%%%%%%%
\section{Recovering the current value of cosmological constant}
\label{cinque}
%%%%%%%%%%%%%%%%%%%%%%%%%%%%%%%%%%%%%
Another important quantum-cosmological relation is the Zeldovich relation for cosmological constant.
In Ref. \cite{zeldovich}, Zeldovich, inspired by the Dirac and Eddington works on  the large numbers coincidence  \cite{dirac1,dirac2,dirac3,dirac4,eddington}, was able to relate the cosmological constant  to the mass of the proton.
Basically the ratio
\begin{equation}
 \frac{R_u}{\lambdabar_p}\,,\label{24}
 \end{equation}
 of the universe radius $R_u$ to the Compton wavelength of the proton $\displaystyle{\lambdabar_p=\frac{\hbar}{M_p c}}$,  the ratio of the age of the universe $\displaystyle{T_u=\frac{R_u}{c}}$ to the characteristic time for the light to cross the proton $\displaystyle{{T_p}=\frac{\lambdabar}{c}}$ are of the order $10^{40\div42}$.
On the other hand, a dimensionless quantity, characterizing the gravitational interaction for the proton, was observed to be  almost of the same order of magnitude
\begin{equation}
\frac{\hbar c}{GM^2_p}\simeq 10^{38}\,.\label{25}
\end{equation}
Without entering into the question of variability of the first two ratios with the expansion of the universe or the  question of  $G$ variability  \cite{zeldovich}, Dirac strongly believed that the numerical agreement between the ratios given in the above  Eqs.(\ref{24}) and (\ref{25}) was not accidental but had some deep meaning (Coincidence Principle).
Equating Eq.(\ref{24}) with (\ref{25}), an interesting formula can be obtained relating again the radius of the universe $R_u$ to the Compton wavelength of the proton:

%Basically he notice that the ratio $\displaystyle{\frac{R_u}{\frac{\hbar}{m_p c}}\simeq 10^{42}}$ and consequently $\displaystyle{\frac{T_u}{\frac{\hbar}{m_p c^2}}\simeq 10^{42}}$ where $\displaystyle{T_u=\frac{R_u}{c}}$. Furthermore it was noticed that also another interesting relation given by $\displaystyle{\frac{\hbar c}{G m_p^2}\simeq 10^{38}}$ was not casually coinciding numerically with the previus ones, but it was strongly believed that this coincidence was a fundamental one.
%Equating them a interesting relation wad obtained relating the radius of the Universe to the Compton wavelength of the proton:
\begin{equation}
R_u=\left(\frac{\hbar }{M_p c}\right)^3 \frac{c^3}{G\hbar} \,.
\label{26}
\end{equation}
Assuming the presence of  cosmological constant $\Lambda$,  the Coincidence Principle can be reinforced replacing the universe radius $R_u$ by the quantity $\Lambda^{-\frac{1}{2}}$ which has the same  dimensions. Zeldovich obtained  \cite{zeldovich}:   

\begin{equation}
\Lambda=\left(\frac{\hbar}{Gc^3}\right)^2\left(\frac{M_p c}{\hbar}\right)^6\end{equation}
which is an outstanding expression for the cosmological constant even though it is six orders of magnitude greater than its current value. It is possible to recover immediately the correct numerical value for $\Lambda$  adopting our approach.
In fact, using Eq. ({\ref{21}}) for $R_u$ and $\displaystyle{\sqrt{n^u_p}=10^2\left(\frac{M_{BH}}{M_p}\right)^2}$ from Eq.(\ref{17}), we get
\begin{equation}
R_u=10^3\left(\frac{\hbar}{M_p c}\right)^3\frac{1}{l_{Planck}^2}\,,
\end{equation}
where the Planck length $\displaystyle{l_{Planck}=\sqrt{\frac{\hbar G}{c^3}}}$ has been adopted. Then, we obtain 
\begin{eqnarray}
\Lambda &=&\frac{1}{10^6}\left(\frac{l_{Planck}^4}{\lambdabar_p^6}\right)=\nonumber\\ &&=\frac{1}{10^6}\left(\frac{\hbar G}{c^3}\right)^2 \left(\frac{M_p c}{\hbar}\right)^6\simeq 10^{-60}cm^{-2}\,,
\end{eqnarray}
which agrees with the order of  magnitude of the observed cosmological constant. In fact, evaluating the corresponding energy density $\displaystyle{\rho_\Lambda=\frac{c^2\Lambda}{8\pi G}}$, one obtains

\begin{equation}
\rho_\Lambda=\frac{1}{10^6}\left(\frac{GM^2_p}{8\pi \hbar c}\right)\left(\frac{M_p c}{\hbar}\right)^3\simeq10^{-29}\frac{g}{cm^3}
\end{equation}
There is another conceptually interesting way  to write  the above equation. It consists  in  introducing the number of protons in the universe, according to
  
  \begin{equation}
  \frac{G M_p^2}{\hbar c}=\left(\frac{M_p}{M_{Planck}}\right)^2=\frac{10^2}{\sqrt{n_p^u}}\,,
  \end{equation}
 derived from  Eq. (\ref{17}). Substituting into the previous expression, we obtain
  
  \begin{eqnarray}
 \rho_\Lambda\simeq \frac{4}{\sqrt{n_p^u}}\left(\frac{M_p}{\frac{4}{3}\pi R_p^3}\right),
   \end{eqnarray}
where $R_p$ is the radius of the proton according to Eq. (\ref{21}). 
Notice that  the  fluctuation factor  $\left(\sqrt{n_p^u}\right)^{-1}$ appears in the above equation and it has a relevant role in the statistical hypothesis \cite{scott1,scott2}. We can multiply and divide by the factor $n_p^u$ obtaining

  \begin{eqnarray}
 \rho_\Lambda\simeq 4\left(\frac{M_u}{\frac{4}{3}\pi R_u^3}\right)\simeq 4 \rho_m
  \end{eqnarray}
  where $\rho_m$ is the matter density of the universe. In this way, the so-called {\it Coincidence Problem}, consisting in the fact that the today observed density of dark energy and dark matter are unnaturally comparable in order of magnitude,  could be naturally addressed. It is important to stress  that also here the scaling relation
   (\ref{26}) has been used.

%\frac{1}{8\pi} \frac{1}{10}\left(\frac{m_p}{\sqrt{n_p^u}}\right)\left[\frac{1}{10\lambdabar_p}\right]^3 
 %%%%%%%%%%%%%%
\section{Conclusions}
\label{sei}
%%%%%%%%%%%%%%%%%%%%%%%%%%%%%

In this paper we have shown  that the physical sizes of self-gravitating  astrophysical structures,  roughly described by Eq. (\ref{20}), are naturally "imprinted"   at the very early stages of the Universe, that is at the time when quantum fluctuations play a crucial role. The quantization relation (\ref{9}) rules, in principle,  all the self-gravitating systems up to the whole universe.
In other words, such a relation  provides  a straightforward generalization of the Eddington-Weinberg relation starting from the Dirac quantization.
Furthermore, also the Zeldovich relation  between the cosmological constant $\Lambda$ and the  Compton wavelength of the proton can be easily recovered in our approach.  In this case, the so-called Coincidence Problem is naturally  addressed. Finally, being the interactions and the granular components  that rule the self-gravitating structures, dark matter is not necessary as further ingredient to build up and stabilize astrophysical systems. 
A final consideration is in order at this point. The presented results are far to be the definite answer to the problem of connecting quantum to cosmological scales, however they could be useful indications in view to address the problems of  large scale structure and  dynamics of self-gravitating astrophysical systems.


\begin{thebibliography}{99}
  
\bibitem{dirac1}
P.M.A. Dirac, {\it Proc. Roy. Soc.} A {\bf 133}, 60, (1931).

\bibitem{dirac2}
P.A.M. Dirac, {\it Proc. Roy Soc.} A {\bf 165}, 199, (1938). 

\bibitem{dirac3}
P.A.M. Dirac, {\it Nature} {\bf139} 323, (1937).

\bibitem{dirac4}
P.A.M. Dirac.{\it Proc. Roy. Soc.}  A {\bf 338}, 439, (1974).

\bibitem{eddington}  
A.S. Eddington, {\it Proc. of the Cambridge Philosophical Society}, {\bf 27}, 15 (1931).

\bibitem{weinberg1}
S. Weinberg, {\it Gravitation and Cosmology}, Wiley, New York (1972).

\bibitem{weinberg2}
 S.  Weinberg, {\it  Rev. Mod. Phys.}  {\bf 61}, 1 (1989).
 
 \bibitem{demartino}
 S. Capozziello, S. De Martino, S. De Siena, F. Illuminati, {\it Mod. Phys. Lett.} A {\bf  15}, 1063 (2000).
 
 
\bibitem{halliwell} 
J.J.Halliwell
 {\it Proceedings of the 1990 Jerusalem Winter School on Quantum Cosmology and Baby Universes},  Ed. S.Coleman, J.B.Hartle, T.Piran and S.Weinberg 
 (World Scientific, Singapore, 1991).
 

\bibitem{SGF}
S. Capozziello, G. Cristofano, M. De Laurentis, {\it  Eur. Phys. J. C} {\bf 69}, 293, (2010).
  

\bibitem{kallosh} R. Kallosh, A. D. Linde, T. Ort\'in, A. W. Van Proeyen, {\it Phys. Rev. D} {\bf 46}, 5278, (1992).

\bibitem{fubini}
S. Fubini, {\it Mod.Phys.Lett.}   {\bf A 6}, 347 (1991).

\bibitem{Gthooft}
G. 't Hooft, {\it Nucl. Phys.}  {\bf 138}, 1 (1978).

\bibitem{Ferrara}
S. Ferrara, K. Hayakawa, A. Marrani, {\it Fortsch. Phys. } {\bf 56}, 993 (2008).

\bibitem{scott1}
S. Capozziello and  S. Funkhouser,
{\it Mod. Phys. Lett.} {\bf  A 24}, 1121 (2009).

\bibitem{scott2}
S. Capozziello and  S. Funkhouser,
{\it Mod. Phys. Lett.} {\bf  A 24}, 1743 (2009).

\bibitem{calogero}
F.  Calogero,  {\it Phys.  Lett.}  {A 228}, 335 (1997).

\bibitem{roy}
S. Roy, {\it Statistical Geometry and Applications to Microphysics and Cosmology}
Kluwer  (1998).

\bibitem{farabook}
S. Capozziello and V. Faraoni, {\it Beyond Einstein Gravity: A Survey of Gravitational Theories for Cosmology and Astrophysics}, Springer, New York (2011).

\bibitem{Tula}
S. Mendoza, X. Hernandez, J. C. Hidalgo,  and T. Bernal, {\it Mon. Not. R.Astr.} {\bf 411}, 226 (2011).

\bibitem{zeldovich}
Ya. B. Zeldovich, {\it Pisma ZhETF} {\bf 6}, 883 (1967).

\end{thebibliography}
\end{document}